	\documentclass[final,3p,twocolumn]{elsarticle}

	\usepackage[english]{babel}
	\usepackage{amssymb}
	\usepackage{amsmath}
	\usepackage{slashed}
	\usepackage{graphicx}
	\usepackage{natbib}
	\usepackage{setspace}
	\usepackage{subfigure}
	\usepackage{appendix}
	\usepackage[pagewise,columnwise]{lineno}

\journal{Physics Letters A}

\graphicspath{{Figures/}}

\begin{document}

\begin{frontmatter}

\title{Preparation of non-ergodic states in quantum spin chains}

 \author[IFGW]{A. O. Garc\'{i}a Rodr\'{i}guez}
 \author[IFGW]{G. G. Cabrera}
 \address[IFGW]{Instituto de F\'isica ``Gleb Wataghin", Universidade Estadual de Campinas, UNICAMP, 13083-859, Campinas, SP, Brazil}

\begin{abstract}

We test the time evolution of quite general initial states in a model that is exactly solvable, \emph{i.e.} a semi-infinite $XY$ spin chain with an impurity at the boundary. The dynamics is portrayed through the observation of the site magnetization along the chain, focusing on the long-time behavior of the magnetization, which is estimated using the stationary phase method. Localized states are split off from the continuum for some regions of the impurity parameter space. Bound states are essential for the non-ergodic behavior reported here. When two impurity states exist, the quantum interference between them leads to magnetization oscillations which settle over very long times with the absence of damping. The frequency of the remanent oscillation is recognized as being the Rabi frequency of the localized levels.
\end{abstract}



\end{frontmatter}

\section{Introduction}
The time evolution of non-equilibrium states in many-particle quantum systems continues to stay as an open issue. As a quantum system, we mean here an interacting and closed quantum system with infinitely many degrees of freedom, in the absence of decoherence or dissipation effects. In the thermodynamic limit, one can argue that parts of the system act as `reservoirs' for other parts, to assure at long times, the relaxation to equilibrium of arbitrary (non-equilibrium) initial states. This property is recognize to be a consequence of \emph{quantum ergodicity}, a concept that was addressed by von Neumann in 1929 \cite{vonneumann}, and is most relevant for the foundations of Statistical Mechanics. In order to better understand the above assumption, one can test non-equilibrium properties in models that are exactly solvable, and eventually find evidence of non-ergodic behavior~\cite{noneq}. Spin chains are considered as useful examples of non trivial quantum interacting systems that may fulfill the above requirements. The relaxation of inhomogeneous initial states has been studied in a number of articles \cite{berimJETP,berimLowT,berimPhysA,berimPRB,karevski,anomalous}, using spin chains with cyclic boundary conditions as prototype systems. In the thermodynamic limit, the relaxation mechanism is driven exclusively by quantum fluctuations and interference effects. A slow relaxation is obtained from contributions of stationary points, where the interference is constructive. Degenerate stationary points present anomalous behavior, reducing even more the damping of the initial excitations as a function of the parameters of the model. But in spite of being very slow, the relaxation process leads those systems to a final homogeneous and stationary state at asymptotically very long times \cite{berimPRB}, a situation that we call \emph{equilibrium state}, in agreement with the ergodic hypothesis. All the systems referred above are amenable of exact analytical solutions, they all have full translational symmetry and are solved with periodic boundary conditions. In the present contribution, we also study the dynamics of a spin chain, but we have broken the translational symmetry, considering a semi-infinite system with an open boundary, where we have attached an impurity. The impurity is characterized by two parameters, its magnetic moment and the coupling with the chain, which may be different from those in the bulk of the system. In the thermodynamic limit (semi-infinite chain), the spectrum has a continuous branch, that we call `the band', and depending on values of the impurity parameters, localized states may be split off from the band. This latter fact is determinant for the phenomena we are dealing with here, that is the preparation of an inhomogeneous (out-of-equilibrium) state that does not relax to equilibrium at asymptotically very long times.
For definiteness, in this work we study the quantum dynamics of a semi-infinite \emph{XY} spin chain with an impurity. This problem can be solved exactly, using the diagonalization method of Ref.~\cite{tsukernik}. In the above approach, the Jordan-Wigner transformation maps the spin Hamiltonian into a free-fermion problem, which can be solved in closed form through a unitary transformation. Due to the absence of translational symmetry, one has to rely in calculations performed in the site representation (real space). Assuming a quite general inhomogeneous initial state for the spin chain, exact results are obtained for the long-time dependence of the magnetization at the different sites of the chain. The stationary phase method has been employed to estimate this long-time behavior. Non-ergodic time evolution is observed, when two localized impurity levels are split off from the band (one above and one below the band edges). For an arbitrary inhomogeneous initial state, the time evolution displays strong oscillations in the site magnetizations. Most of the oscillations are quickly damped by destructive interference, stationary point contributions relax more slowly, as in the examples previously mentioned in Ref.~\cite{berimPhysA,berimPRB,anomalous}, and at asymptotically very long times it only remains an undamped oscillation, with a frequency that is identified as the Rabi frequency of the two localized levels. If the inhomogeneity of the initial state is chosen near the open boundary, the amplitude of the oscillation is paramount at the first sites of the chain. It decreases in magnitude when going to the bulk of the sample, but it is not damped in time.
The phenomenon reported here is similar to the one observed experimentally in $1-D$ Bose gases confined by light traps \cite{cradle}. After the preparation of out-of-equilibrium initial states, non-ergodic behavior is observed, with the absence of damping after thousands of atom collisions. Concerning spin systems, proper experimental setups are nowadays available. Thanks to the progress of experimental techniques, it is possible to create quantum spin chains~\cite{brune}, for both, ferromagnetic and antiferromagnetic couplings ~\cite{gambardella1,hirjibehedin}. Devices which are candidates for building a quantum computer, such as Josephson junction arrays, mimic the physics of quantum spin chains~\cite{makhlin}. Inhomogeneous initial excitations can be produced experimentally in those systems, by locally manipulating gate voltages, tunneling barriers or magnetic fields.

In the next Section of this paper, we discuss the model used in the calculation, along with the method of solution. After this, we schematically describe the time evolution for a particular (but typical) inhomogeneous initial state. Results for the long-time behavior of the magnetization are displayed in the last Section, where we state the final conclusions.
\section{The model}
\begin{figure}[h!]
\centering
\includegraphics[width=\linewidth]{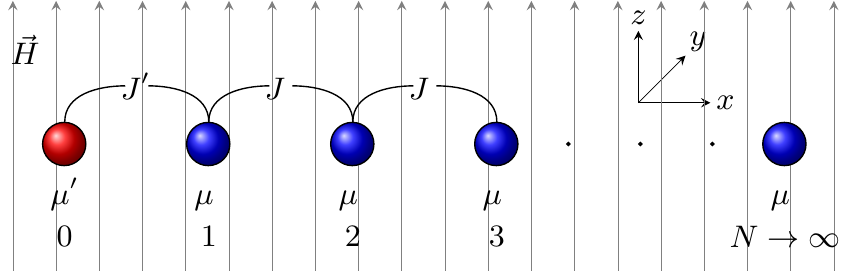}
\caption{(Color online) Semi-infinite quantum spin chain with an impurity `atom' at site \(n=0\). A uniform external magnetic field $H$ is applied along the \(z\)-axis .}
\label{chain}
\end{figure}

We consider a semi-infinite quantum spin chain with an impurity at one end of the chain, as depicted in Fig.~\ref{chain}.

\begin{figure}[h!]
\centering
\includegraphics[width=\linewidth]{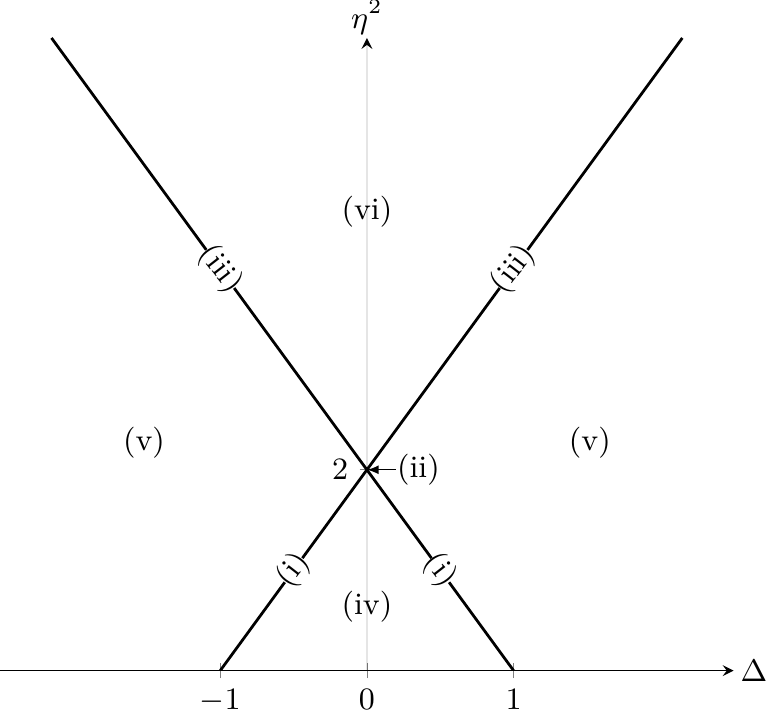}
\caption{Characteristic regions for the presence of localized levels in the spectrum. See explanations in the main text.}
\label{regs}
\end{figure}

The figure is self-explanatory, $\mu$ is the magnetic moment in the bulk of the chain, where spins are coupled with exchange $J$. The corresponding quantities for the impurity are $(\mu',J')$. For convenience, to follow the solution of Ref.~\cite{tsukernik}, we define the parameters:
\begin{equation}
\eta=J'/J\qquad\Delta=-\left(\mu'-\mu\right)H/J~.
\end{equation}
We model our spin chain using a \emph{XY}-Hamiltonian for spin $s=1/2$ with nearest neighbor interactions in the presence of a uniform transverse magnetic field $H$:
\begin{equation}
\begin{aligned}
\mathcal H\!=\!&-J'\!\!\left(S_0^{\mathstrut x} S_1^{\mathstrut x}\!+\!S_0^{\mathstrut y} S_1^{\mathstrut y}\right)\!-\!J\!\!\sum_{n=1}^{N-1}\!\!\left(S_{n\vphantom1}^{\mathstrut x} S_{n+1}^{\mathstrut x}\!+\!S_{n\vphantom1}^{\mathstrut y} S_{n+1}^{\mathstrut y}\right)\\
&-\mu' H S_0^{\mathstrut z}-\mu H\sum_{n=1}^N S_{n\vphantom1}^{\mathstrut z},
\end{aligned}
\label{ham}
\end{equation}
and we consider the case of a semi-infinite spin chain (\(N\to\infty\)) for which the hamiltonian \eqref{ham} can be exactly diagonalized~\cite{tsukernik}. The infinite number of degrees of freedom avoids revivals of the initial quantum state, which may be present in finite size systems~\cite{revival}. For details of the solution, we refer the reader to the paper~\cite{tsukernik}. For completeness, we briefly summarize the procedure. We firstly use the Jordan-Wigner transformation, to express spin-1/2 operators in terms of fermion annihilation and creation operators \(c_n^{\mathstrut}\) and \(c_n^\dag\):
\begin{equation}
S_0^+=c_0^\dag,\qquad S_n^+=c_n^\dag\prod_{l=0}^{n-1}\left(1-2c_{l}^\dag c_{l}\right),\quad n\geqslant1,
\label{jw}
\end{equation}
where $S_n^+$ are the spin ladder operators. The resulting fermion Hamiltonian can be diagonalized by a unitary transformation that preserves the anti-commuting properties:
\begin{equation}
c_n^{\mathstrut}=\sum_{\lambda=1} u_{n\vphantom1}^{(\lambda)} b_\lambda^{\mathstrut},\quad n\geqslant0,
\label{transfunit}
\end{equation}
where the coefficients \(u_{n\vphantom1}^{(\lambda)}\) (wave functions) are derived from the Heisenberg equations of motion for the quasiparticle operators $b_\lambda$, proviso that the Hamiltonian acquires the quasi-particle form:
\begin{equation}
\mathcal H=\sum_{\lambda=1} \epsilon_\lambda~ b_\lambda^\dag\ b_\lambda ,
\end{equation}
with $\lambda$ being the `good quantum number' characterizing the energy spectrum. As in any eigenvalue problem, the equations of motion lead to a set of coupled linear equations in finite difference for the wave functions $u_{n}^{\lambda}$ (which is infinite in the thermodynamics limit). This set of equations is solved by recurrence methods, yielding the eigenvalues ${\epsilon_\lambda}$ as well as the corresponding wave functions $u_{n}^{\lambda}$~\cite{tsukernik}. As a general rule, one gets a continuous spectrum corresponding to the bulk of the chain. Depending on values of the impurity parameters, one can split off two, one, or none localized levels from the band. The different regions of solution in the impurity parameter space are displayed in Fig.~\ref{regs}, with coordinates $\Delta$ and $\eta^2$.
The symbols \emph{(i)} and\emph{(iii)} are used to denote borders, and \emph{(ii)} is the crossing at $\Delta=0,~\eta^2=2$. The four characteristic regions of solutions are labeled by \emph{(iv)}, \emph{(v)}, with $\Delta>0$, \emph{(v)}, with $\Delta<0$, and \emph{(vi)}. There are no bound states in region \emph{(iv)}, while there is one bound level in regions \emph{(v)}; with energy lower than the band, for $\Delta<0$, and higher than the band for $\Delta>0$. In region \emph{(vi)} there are two localized levels, one above and one below the continuous spectrum.
With this final remark, we end the description of the exact solution obtained in Ref.~\cite{tsukernik}. With the above solution in hand, we turn to the calculations of the dynamical properties of the chain, when an inhomogeneous state is prepared as initial state. The detailed calculations for the four regions, the borders and the crossing will be published in a separate contribution~\cite{next}. Here, we will focus in a few relevant examples.

\section{Time evolution}
We assume an initial state for the chain in which each `atom' is in one of the two eigenstates of the \(z\)-component of spin, \(\left|1/2\right\rangle\) (spin up) or \(\left|-1/2\right\rangle\) (spin down). The system initial state then reads
\begin{equation}
\left|\Psi(0)\right\rangle=\left|m_z^{(0)}(0), m_z^{(1)}(0), \dots, m_z^{(N)}(0)\right\rangle,
\label{estin}
\end{equation}
with
\begin{equation}
m_z^{(n)}(0)=\pm1/2,\quad n\geqslant0.
\label{valsestin}
\end{equation}

The above is an eigenstate of $S^z$, the $z$-component of the total spin, but it is not an eigenstate of Hamiltonian (\ref{ham}), so it displays a non trivial dynamics. To describe the latter, we calculate the \(z\)-axis magnetization for each site \(n\) at an arbitrary instant of time \(t\). Denoting this quantity by \(\left\langle S_{n\vphantom1}^{\mathstrut z}\right\rangle_{\!\!t}\), we have
\begin{equation}
\left\langle S_{n\vphantom1}^{\mathstrut z}\right\rangle_{\!\!t}=\left\langle\Psi(t)\,\middle|\,S_{n\vphantom1}^{\mathstrut z}\,\middle|\,\Psi(t)\right\rangle,
\label{defmagn}
\end{equation}
where \(\left|\Psi(t)\right\rangle=e^{-\frac i\hbar\mathcal H t}\left|\Psi(0)\right\rangle\) is the system state at time \(t\). The Schr\"{o}dinger expectation value $\left\langle S_{n\vphantom1}^{\mathstrut z}\right\rangle$ can be written in the Heisenberg picture, with
\begin{equation}
\left\langle S_{n\vphantom1}^{\mathstrut z}\right\rangle_{\!\!t}=\left\langle S_{n\vphantom1}^{\mathstrut z}(t)\right\rangle_{\!\!0},\quad S_{n\vphantom1}^{\mathstrut z}(t)=e^{\frac i\hbar\mathcal H t}\,S_{n\vphantom1}^{\mathstrut z}\,e^{-\frac i\hbar\mathcal H t}~,
\label{defmagnestin}
\end{equation}
where the average in the latter is taken with the initial state.
The Heisenberg operator \(S_{n\vphantom1}^{\mathstrut z}(t)\) can be expressed in terms of the quasiparticle operators \(b_\lambda^{\mathstrut}\) and \(b_\lambda^\dag\), whose trivial time evolution
\begin{equation}
b_\lambda^{\mathstrut}(t)=b_\lambda^{\mathstrut} e^{-\frac i\hbar\epsilon_\lambda^{\mathstrut} t},\quad b_\lambda^\dag(t)=b_\lambda^\dag e^{\frac i\hbar\epsilon_\lambda^{\mathstrut} t},
\label{bsHeins}
\end{equation}
allows us to obtain the explicit time dependence of the magnetization. Subsequently, using the result for arbitrary \(t\), we address the question of obtaining the magnetization for very large values of time, i.e., the long-time tails. In the thermodynamic limit ($N\rightarrow\infty$), this is done using the stationary phase method~\cite{SPM}, to estimate the dominant behavior at asymptotically long times. Once the initial state is given in the form (\ref{estin}), the site magnetization can be written as\footnote{Several equivalent exact expressions for the site-magnetization can be obtained~\cite{next}, and they are more or less convenient, depending on the structure of the initial state. Relation (\ref{3rdresger}) given in the text is the most compact one.}:
\begin{equation}
\left\langle S_{n\vphantom1}^{\mathstrut z}\right\rangle_{\!\!t}=\sum_{m=0}^Nm_z^{(m)}(0)S_{n,m}(t),\quad n\geqslant0~,
\label{3rdresger}
\end{equation}
where all the interference effects are contained in the factor $S_{n,m}(t)$, which is written as
\begin{equation}
S_{n,m}(t)=\left|\sum_{\lambda=1}~u_{n\vphantom1}^{(\lambda)} {u_{m\vphantom1}^{(\lambda)}}^*e^{-\frac i\hbar\epsilon_\lambda^{\mathstrut} t}\right|^2~,
\label{Snm}
\end{equation}
with \{$u_{m}^{\lambda}$\} being the wave functions obtained through the diagonalization process (see Eq. (\ref{transfunit})). In the general case, the sum (\ref{Snm}) includes the continuous as well as the discrete spectra. Exact asymptotic series can be obtained for the long-time behavior of $S_{n,m}(t)$. In general, dominant terms of the series come from contributions of stationary points and they show a very slow relaxation, in the form of a power law $t^{-\alpha}$, with $\alpha\geq1$~\cite{next}. There are remarkable exceptions that we want to discuss here. To better illustrate the above phenomena, we will choose the specific initial state given below:
\begin{equation}
\left|\Psi(0)\right\rangle=\left|\uparrow\right\rangle_{\!0}\left|\downarrow\right\rangle_{\!1}\left|\uparrow\right\rangle_{\!2}\left|\uparrow\right\rangle_{\!3}\dots\left|\uparrow\right\rangle_{\!N}~,
\label{specific}
\end{equation}
where all the spins are pointing up, except the second which is coupled to the impurity and pointing down. For this state, the site-magnetization acquires the very simple form
\begin{equation}
\left\langle S_{n\vphantom1}^{\mathstrut z}\right\rangle_{\!\!t}=\frac12-S_{n,1}(t),\quad n\geqslant0~.
\label{magnestinesp}
\end{equation}
Results for the time evolution of this state are shown in the next Section, where we also give the final comments.

\section{Results and discussion}
We will present results and figures corresponding to time evolution for values of parameters in regions (\emph{iv}), (\emph{v}) and (\emph{vi}) of Fig.~\ref{regs}, when the initial state is prepared in the form given by (\ref{specific}).

\subsection{The band}
\begin{figure}[h!]
\includegraphics[width=\linewidth]{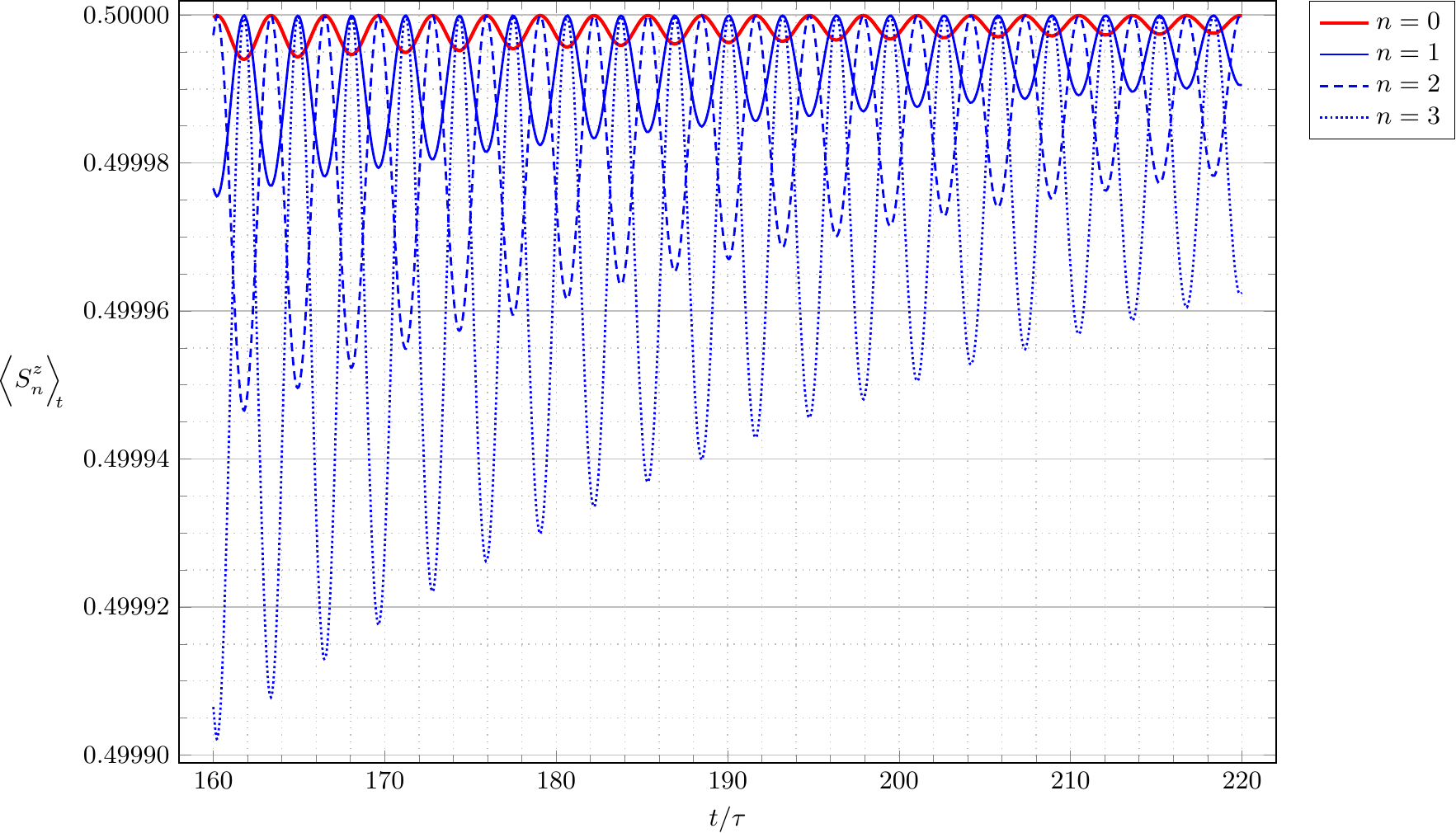}
\caption{(Color online) Long-time behavior of the magnetization at sites $n=0,1,2,3$, for parameters in region (\emph{iv}) of Fig.~\ref{regs}. The examples shown correspond to the values \(\Delta=0\), \(\eta^2=1\). The unit of time is given by \(\tau=\hbar/\left|J\right|\).}
\label{plotreg4n0123}
\end{figure}

For region (\emph{iv}), there are no bound states. The relaxation is driven by states within the continuum, where the interference is in general, destructive. There are two stationary points, which correspond to the edges of the band, which yield a slower evolution in the form of a damped oscillation that relaxes as $ 1/t^3$. This relaxation is due to constructive interference of states in the neighborhood of stationary points. The frequency of the oscillation is the band width, \emph{i.e.} $\omega=2J/\hbar$. In Fig.~\ref{plotreg4n0123}, we display the long-time evolution, for the first sites of the chain ($n=0,1,2,3$).

It is evident from the figure, that at asymptotically long-times, there is only one frequency and all oscillations are damped in time. The magnetization of the extrapolated final state is homogeneous and stationary.
\subsection{One impurity state}\label{one}
In region (\emph{v}), one bound state is split off from the band. There is a symmetry between the two subregions $\Delta>0$ and $\Delta<0$, where the localized state splits from above or below the band, respectively (assuming $J>0$, and if $J$ changes sign, the statement is reversed). For the dynamics, it makes no difference the sign of $J$, and the oscillatory dominant term has two different frequencies superposed. Now the time evolution of the magnetization shows a peculiar property, with oscillations decaying as $1/t^{3/2}$ to a \emph{constant value} which depends on the site. The two frequencies that appear in the long-time behavior are given by the difference between the localized level and the edges of the band:
\begin{equation}
\omega_{\pm}=\frac{\epsilon_i\pm J}{\hbar}~,
\label{frequencies}
\end{equation}
where $\epsilon_i$ is the energy of the bound state. Thus, one gets a `low' and a `high' frequency oscillation in the asymptotic behavior. An example is shown in Fig.~\ref{plotreg5n0123}:

\begin{figure}[h!]
\includegraphics[width=\linewidth]{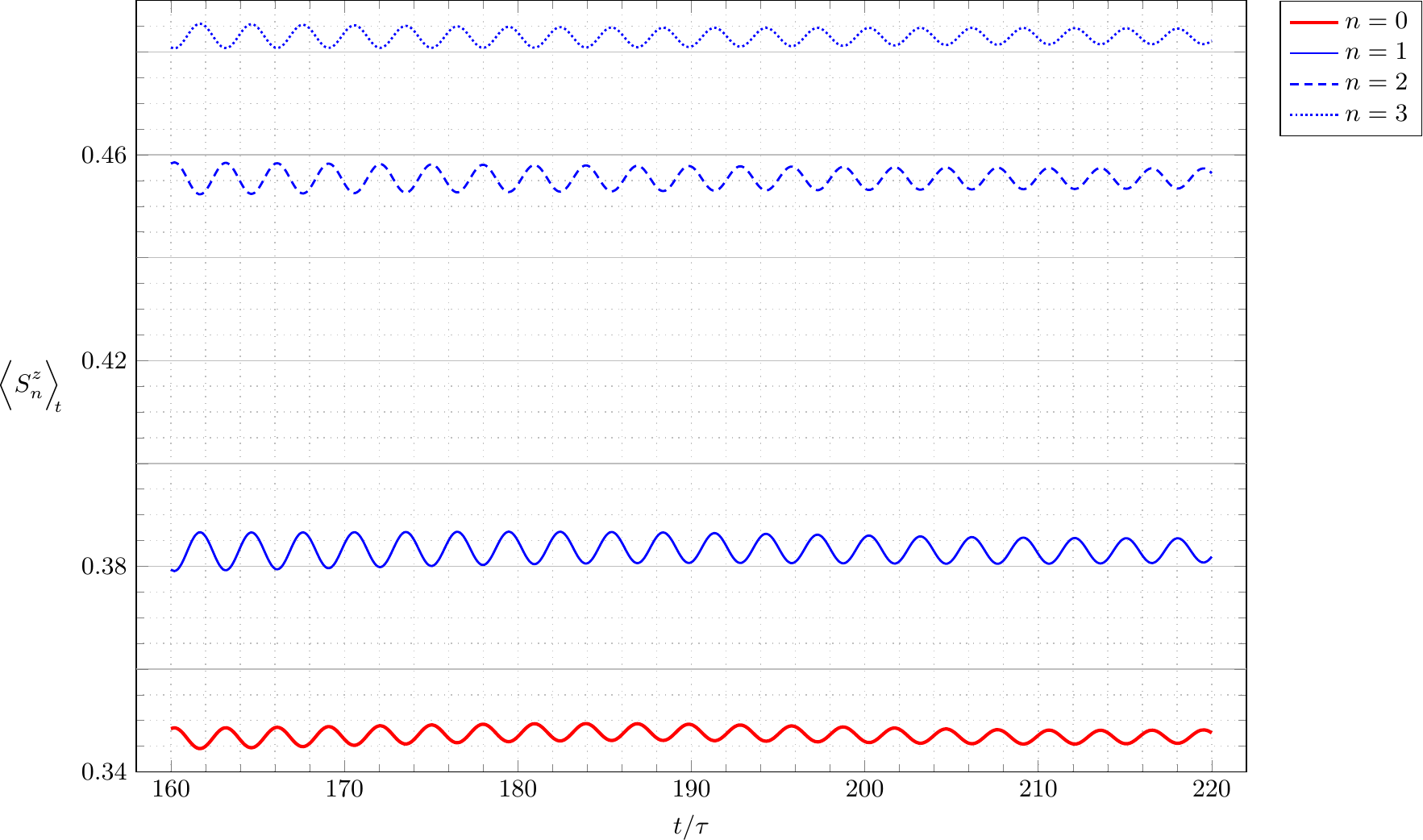}
\caption{(Color online) Long-time behavior of the magnetization at sites $n=0,1,2,3$, for the initial state \eqref{specific}. Values of parameters are \(\left|\Delta\right|=1/2\), \(\eta^2=2\), which correspond to region (\emph{v}). Relaxation is very slow, and due to the time scale, only the oscillation with the `high frequency' is apparent. The unit of time is given by \(\tau=\hbar/\left|J\right|\).}
\label{plotreg5n0123}
\end{figure}
The magnetization extrapolated at $t\rightarrow \infty$ is stationary, but inhomogeneous, showing a non constant profile near the impurity, at asymptotically very long times.

\subsection{Two impurity states}
Region (\emph{vi}) is the characteristic region of parameter space with the presence of two bound states, one above and one below the continuous band. The stationary phase methods yields the dominant terms of the asymptotic series in the form of oscillations with five different frequencies, two `low', two `high', and the highest which corresponds to the Rabi frequency of the two bound states. The `low' and `high' frequencies are similar to those obtained in subsection \ref{one} for the case of one impurity, except that here we have two localized levels. The Rabi frequency is $\omega_{Rabi}=|\epsilon_1-\epsilon_2|/\hbar$~, where $\epsilon_1$ and $\epsilon_2$ are the energies of the bound states. When $\Delta=0$, the levels split symmetrically from the band edges, and the number of frequencies is reduced to three. The `high' and `low' frequency oscillations are damped in time by a factor proportional to $1/t^{3/2}$, as in the previous case. What is \emph{remarkable} in this case is the fact that the Rabi oscillation is not damped in time, and the system is never stationary, nor homogeneous. Examples are shown in Fig.~\ref{plotreg6n0123}:
\begin{figure}[h!]
\includegraphics[width=\linewidth]{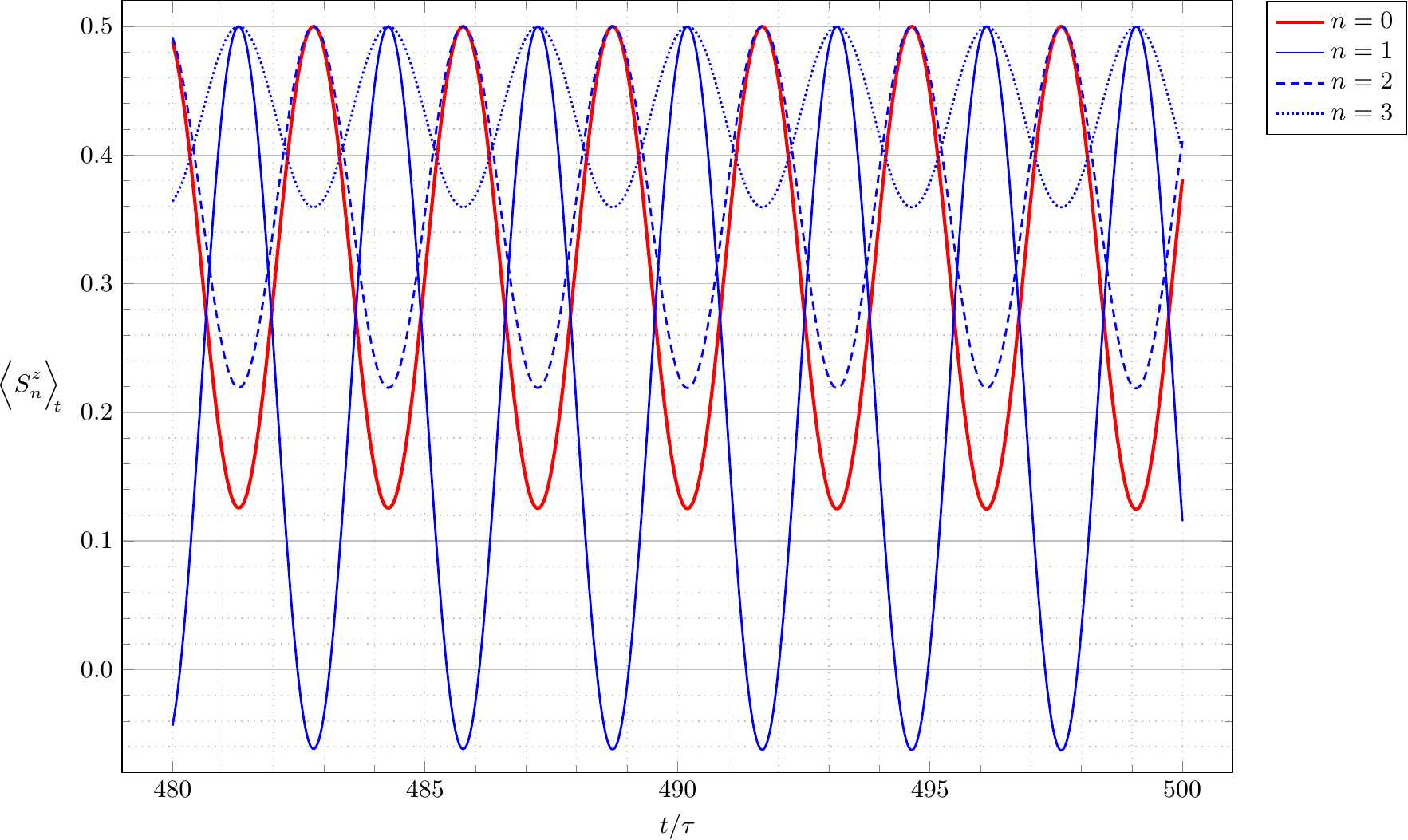}
\caption{(Color online) Long-time behavior of the magnetization at sites $n=0,1,2,3$, for the initial state \eqref{specific}. Values of parameters are \(\Delta=0\), \(\eta^2=3\). Due to the time scale, only the Rabi frequency is visible. Oscillations are undamped. The unit of time is given by \(\tau=\hbar/\left|J\right|\).}
\label{plotreg6n0123}
\end{figure}

\subsection{Final discussions}
In summary, for the same initial state, the non-ergodic behavior appears by changing the impurity parameters in the quantum model. The absence of damping at asymptotically long times, occurs when two localized modes are separated from the continuum, with the impurity parameters in region (\emph{vi}), as shown in Fig.~\ref{regs}. The quantum interference between the two bound states yields magnetization oscillations with the Rabi frequency of the levels. The oscillations settle over very long times with constant amplitude. This effect may have potential applications in quantum computation. In fact, spin chains have been suggested as efficient channels for the short-range transport of information in a quantum computing device~\cite{bose}. Even in finite-size chains, one could find precursory imprints of the undamped oscillation. In addition of the systems mentioned in the Introduction, the effect can also be investigated with polar atoms or molecules trapped in optical lattices, systems that simulate the physics of spin chains. The $XY$ model can be realized by proper combinations of microwave excitations and the many-body interactions induced by the radiation~\cite{simulation}.

\section*{Acknowledgements} \label{Sec-acknowledgements}

The authors are grateful to S\~ao Paulo Research Foundation (\textbf{FAPESP}, Brazil) for financial support through the project No. 2009/53826-8.

\section*{References}

\bibliographystyle{elsarticle-num}
\bibliography{Referencias}

\end{document}